\documentclass[a4paper,11pt]{article}
\pdfoutput=1 

\usepackage{jinstpub} 

\title{\boldmath Planar n-in-n quad module prototypes for the ATLAS ITk upgrade at HL-LHC}

\author[1]{A.~Gisen,\note{Corresponding author.}}
\author{S.~Altenheiner,}
\author{I.~Burmeister,}
\author{C.~G\"o\ss{}ling,}
\author[2]{R.~Klingenberg,\note{Deceased 24 May 2017.}}
\author{K.~Kr\"oninger,}
\author{J.~L\"onker,}
\author{M.~Weers}
\author{and F.~Wizemann}
\affiliation{Experimentelle Physik IV, TU Dortmund\\44221 Dortmund, Germany}

\emailAdd{andreas.gisen@tu-dortmund.de}

\abstract{In order to meet the requirements of the High Luminosity LHC (HL-LHC), it will be necessary to replace the current tracker of the ATLAS experiment. Therefore, a new all-silicon tracking detector is being developed, the so-called Inner Tracker (ITk). The use of quad chip modules is intended in its pixel region. These modules consist of a silicon sensor that forms a unit along with four read-out chips.

The current ATLAS pixel detector consists of planar n-in-n silicon pixel sensors. Similar sensors and four FE-I4 read-out chips were assembled to first prototypes of planar n-in-n quad modules. The main focus of the investigation of these modules was the region between the read-out chips, especially the central area between all four read-out chips. There are special pixel cells placed on the sensor which cover the gap between the read-out chips.

This contribution focuses on the characterization of a non-irradiated device, including important sensor characteristics, charge collection determined with radioactive sources as well as hit efficiency measurements, performed in the laboratory and at testbeams.
In addition, first laboratory results of an irradiated device are presented.
}


\arxivnumber{1708.08266} 

\proceeding{11th International Conference on Position Sensitive Detectors \\
3rd - 8th September 2017 \\
The Open University, Walton Hall, Milton Keynes, U.K.}

\graphicspath{{figures/}} 

\newcommand{\centi}{c}
\newcommand{\milli}{m}
\newcommand{\micro}{\textmu}
\newcommand{\nano}{n}

\newcommand{\meter}{m}
\newcommand{\ampere}{A}
\newcommand{\volt}{V}

\newcommand{\percent}{\%}

\newcommand{\per}[1]{#1$^{-1}$}

\newcommand{\SI}[2]{{#1}\,{#2}} 

\newcommand{\Fig}[1]{\autoref{#1}}

\begin{document}
\maketitle
\flushbottom

\section{ATLAS ITk upgrade}
To cope with the increased luminosity, data rate and radiation damage at the HL-LHC \cite{HL-LHC_PDR}, major upgrades of the \mbox{ATLAS} experiment \cite{ATLAS:Experiment} are foreseen \cite{ATLAS:Phase_II_LoI,ATLAS:Phase_II_Scoping}. 
One upgrade will be the replacement of the current tracking detector.
The proposed new inner tracker (ITk) \cite{ATLAS:TDR-strips} is designed to operate under conditions featuring an increased track density, following from approximately 200 inelastic proton-proton collisions per beam crossing, and a high radiation dose estimated from the expected integrated luminosity of \SI{3000}{\per{fb}} over ten years of operation.

The ITk will be an all-silicon tracker with five pixel and four strip layers in the central region, enclosed by end-caps with six strip disks and a number of pixel rings on each side \cite{ATLAS:TDR-strips}.
Quad chip modules are foreseen in the outer layers and the rings of the pixel region. A quad module consists of four read-out chips which are bump bonded to one silicon sensor. 

\section{n-in-n quad modules}	
\begin{figure}[htbp]
	\centering 
	\includegraphics[width=.65\textwidth]{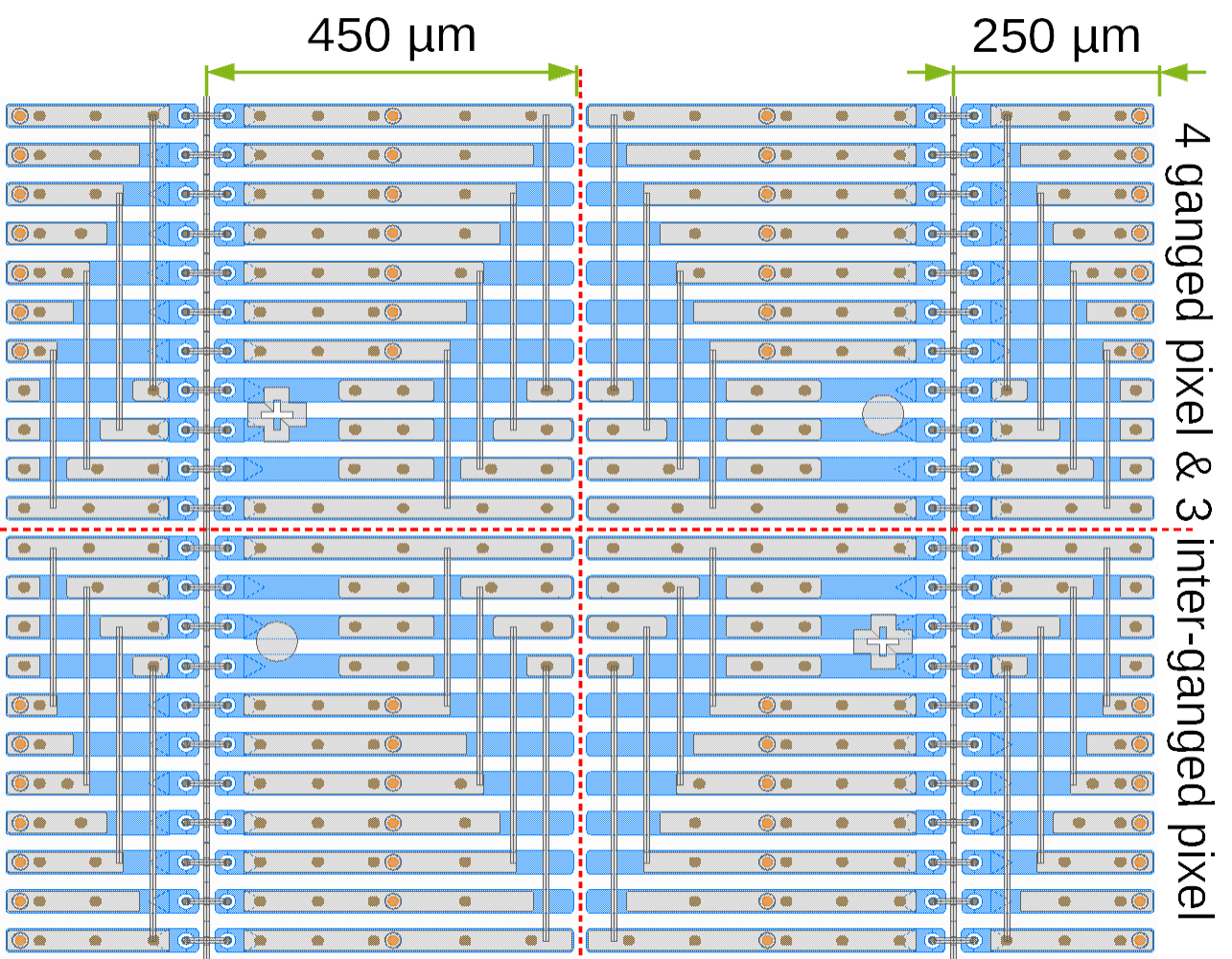}
	\caption{\label{fig:inter-chip} Layout of the sensor's central region with `ganged', `inter-ganged', `long ganged' and `long inter-ganged' pixel cells. The two grey crosses and dots are alignment marks for the flip-chip process. The orange circles represent openings for the bump bond connections.}
\end{figure}

	The two quad module prototypes presented here consist of oxygenated n-doped
	float zone silicon sensors which carry highly n-doped pixel implants with
	moderated p-spray isolation and a p-doped backside.
	The n-bulk thickness is \SI{285}{\micro\meter}.	
	This technology was selected since the current ATLAS pixel detector also consists of such planar n-in-n pixel sensors \cite{Pixel:Electronics_Sensor}.	
	The pixels are connected to a front-end chip with the help of a flip-chip
	process employing tin-lead-bumps.		
	The used FE-I4 front-end chip \cite{GARCIASCIVERES2011S155} was developed for the Insertable B-Layer (IBL) \cite{ATLAS:IBL_TDR,ATLAS:IBL_TDR2} of the ATLAS experiment.
	It has a feature size of \SI{130}{\nano\meter} resulting in \SI{$250\times50$}{\micro\meter$^2$} pixel cells, arranged in 80 columns and 336 rows.
	Four FE-I4 chips with a  bulk thickness of \SI{700}{\micro\meter} are flip-chipped to one sensor.	
	For technological reasons, two chips should not touch each other, which causes gaps in the horizontal and vertical directions.
    To cover the gap in the direction of the long pixel side between two read-out chips, two `long pixels' are extended to a length of \SI{450}{\micro\meter} in every row.    
    To cover the gap in the direction of the short pixel side between two read-out chips, in every column four `ganged pixels' are connected via a metal trace to four pixels without dedicated read-out channels, with three `inter-ganged pixels' in-between. 
    `Inter-ganged pixels' are standard \SI{$250\times50$}{\micro\meter$^2$} pixel cells with a dedicated read-out channel. Because they are enclosed by `ganged pixels', they are treated separately.
    The layout of the sensor's central region is shown in \Fig{fig:inter-chip}, where the combination of these designs leads to special pixel cells like `long-ganged pixels'.
    To reduce inactive sensor area, the first and the last pixel column is shifted completely beneath the guard rings. These pixel columns are referred to as `edge pixels'. 
    
    If all these pixels are taken into account, the sensor consists of $160 \times 680$ individual pixel cells with the external dimensions of \SI{$40.4 \times 17.0$}{\milli\meter$^2$}, resulting in a total sensitive area of \SI{13.736}{\centi\meter$^2$}.
    
    Each quad module assembly is mounted on a PCB. Sensor and front-end pads were wire-bonded to allow calibration and read-out.
    The PCB thickness is \SI{1.5}{\milli\meter}. It has four rectangular openings of \SI{$1.0\times0.8$}{\centi\meter$^2$} in the middle of each read-out chip and an additional opening in the center of the sensor. 
    After investigation in lab and testbeam measurements, one quad was irradiated at CERN-PS IRRAD \cite{Ravotti:2014} in a first step up to a fluence of $5 \times 10^{14}$\,n$_\text{eq}$cm$^{-2}$.
        
\section{IV measurements}
\label{sec:IV}
Current-voltage characteristics (IV curves) reveal important sensor properties and are also a main criterion for quality control.
Certain acceptance criteria must be fulfilled by the sensors in order to be considered for the ITk.
For unirradiated planar pixel sensors, the leakage current at \SI{20}{$^\circ$C} should be less than \SI{0.75}{\micro\ampere\centi\meter$^{-2}$} at the operating bias voltage.
The breakdown voltage should be at least \SI{20}{\volt} higher than the operating voltage.
After irradiation to $2 \times 10^{15}$\,n$_\text{eq}$cm$^{-2}$, for a sensor of \SI{150}{\micro\meter} thickness, the leakage current at \SI{-25}{$^\circ$C} should be less than \SI{20}{\micro\ampere\centi\meter$^{-2}$} at \SI{400}{V}, the breakdown voltage should be higher than \SI{400}{V}.

\begin{figure}[htbp]
   	\begin{minipage}[t]{0.49\textwidth}
   		\centering
   		\includegraphics[width=1.0\textwidth]{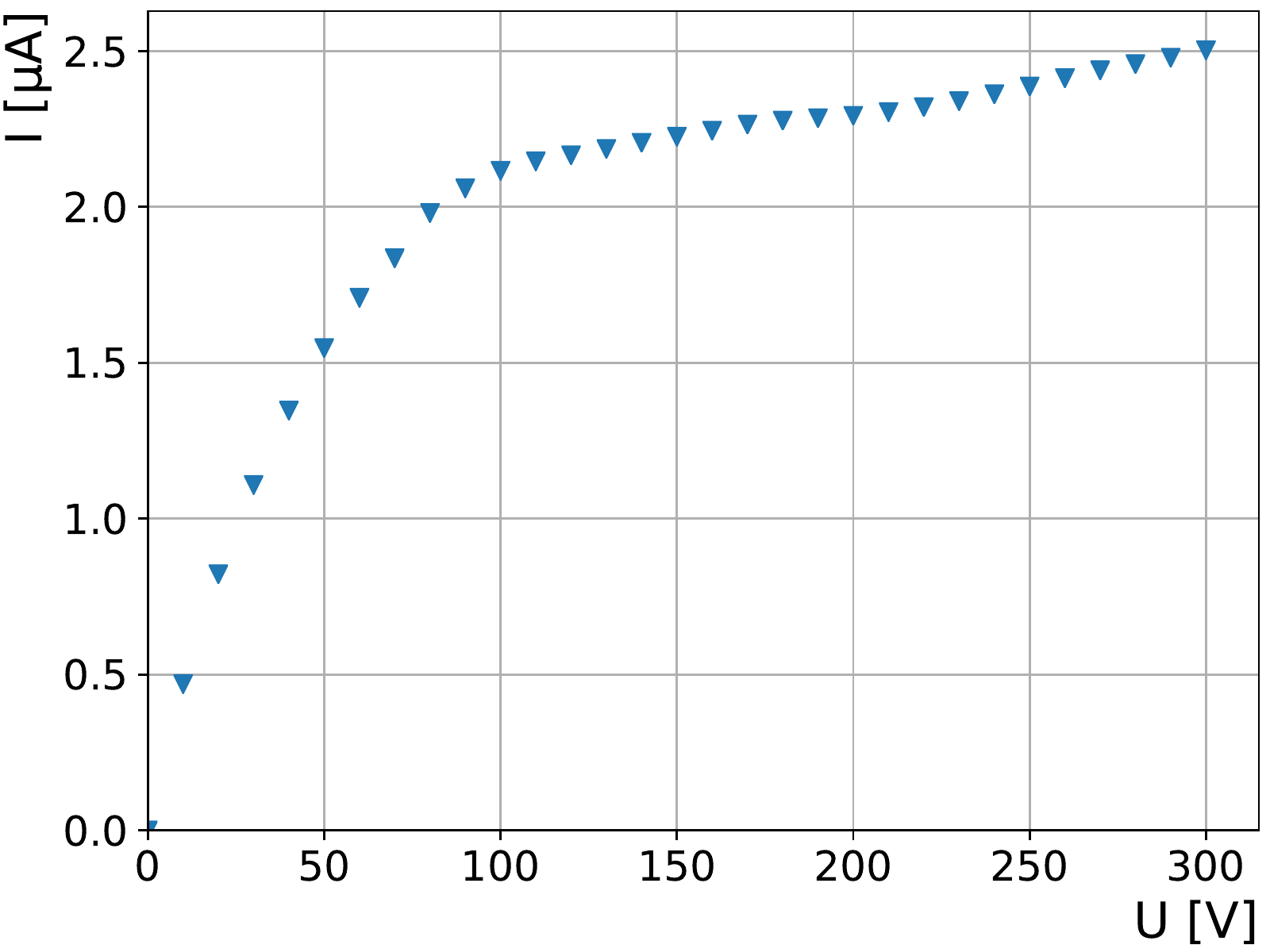}
   		\caption{\label{fig:IV_unirrad} IV curve of the quad module taken before irradiation in a climate chamber at \SI{20}{$^\circ$C}.}
   	\end{minipage}
   	\hfill
   	\begin{minipage}[t]{0.49\textwidth}
   		\centering
	    \includegraphics[width=1.0\textwidth]{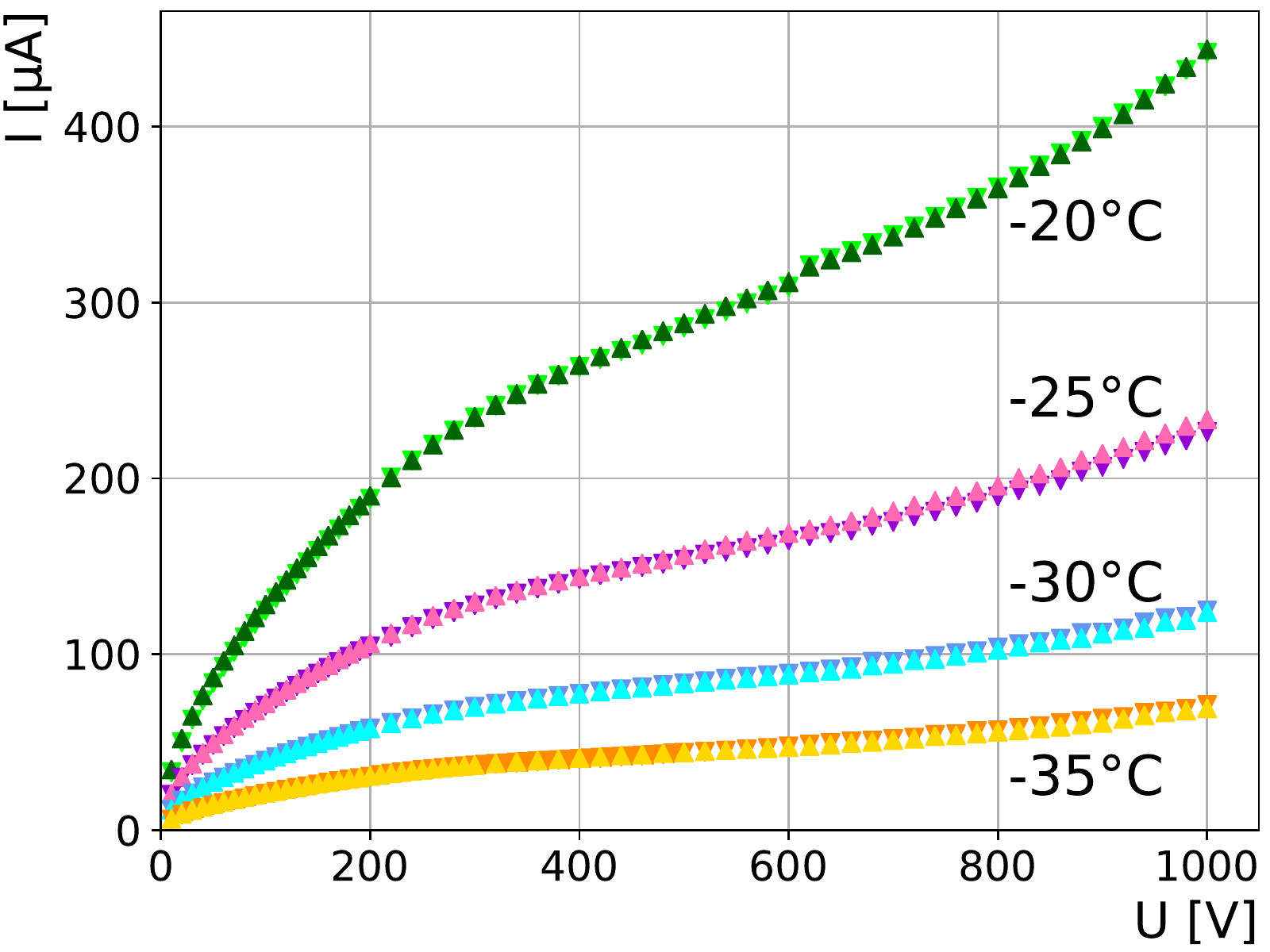}
	    \caption{\label{fig:IV_irrad} IV curves of the quad module taken after irradiation in a climate chamber. Repeated measurements are displayed with a similar colour and a rotated triangle.}
   	\end{minipage}
   	\end{figure}

	Before irradiation, the current was measured up to a maximum voltage of \SI{300}{V} in a climate chamber at \SI{20}{$^\circ$C}. The diode-like curve is shown in \Fig{fig:IV_unirrad}. At \SI{100}{V}, where the plateau starts, a current of \SI{2.1}{\micro\ampere} is measured. The slope is determined to be \SI{1.9}{\nano\ampere\per\volt}. Thus, the acceptance criteria are fulfilled, the current is far below the \SI{10.3}{\micro\ampere} requested for a sensor with an area of \SI{13.7}{\centi\meter$^2$} and no breakdown occurs.
		
	After irradiation, the maximum voltage was increased to \SI{1000}{V}. Measurements were performed at multiple temperatures in a climate chamber. 
	To ensure reproducibility, a second curve was recorded immediately after the first.
	As shown in \Fig{fig:IV_irrad}, again no breakdown occurred. Since only a fluence of $5 \times 10^{14}$\,n$_\text{eq}$cm$^{-2}$ was investigated, no statement about the acceptance criteria can be made.
    The given temperature corresponds to the set temperature of the climate chamber. An offset in sensor temperature caused by self heating is expected but not taken into account.
    
    The humidity inside the climate chamber was controlled and kept low by its circuit, but it was not monitored.

\section{Tuning of front-end chips}
\label{sec:tuning}
Each pixel read-out cell of the FE-I4 chip contains a discriminator with an adjustable threshold.
If a signal in the sensor exceeds this threshold, the time over threshold (ToT) is measured.
This ToT is therefore directly related to the induced charge in the sensor.
The threshold value and the ToT response at an injected reference charge can be adjusted by various DACs. This is necessary to guarantee homogeneous responses under different conditions because the FE-I4 electronics are susceptible to influences such as temperature and radiation. The procedure of adjusting the response of all pixels is called \emph{tuning}.

\begin{figure}[htbp]
\centering 
\includegraphics[width=.49\textwidth]{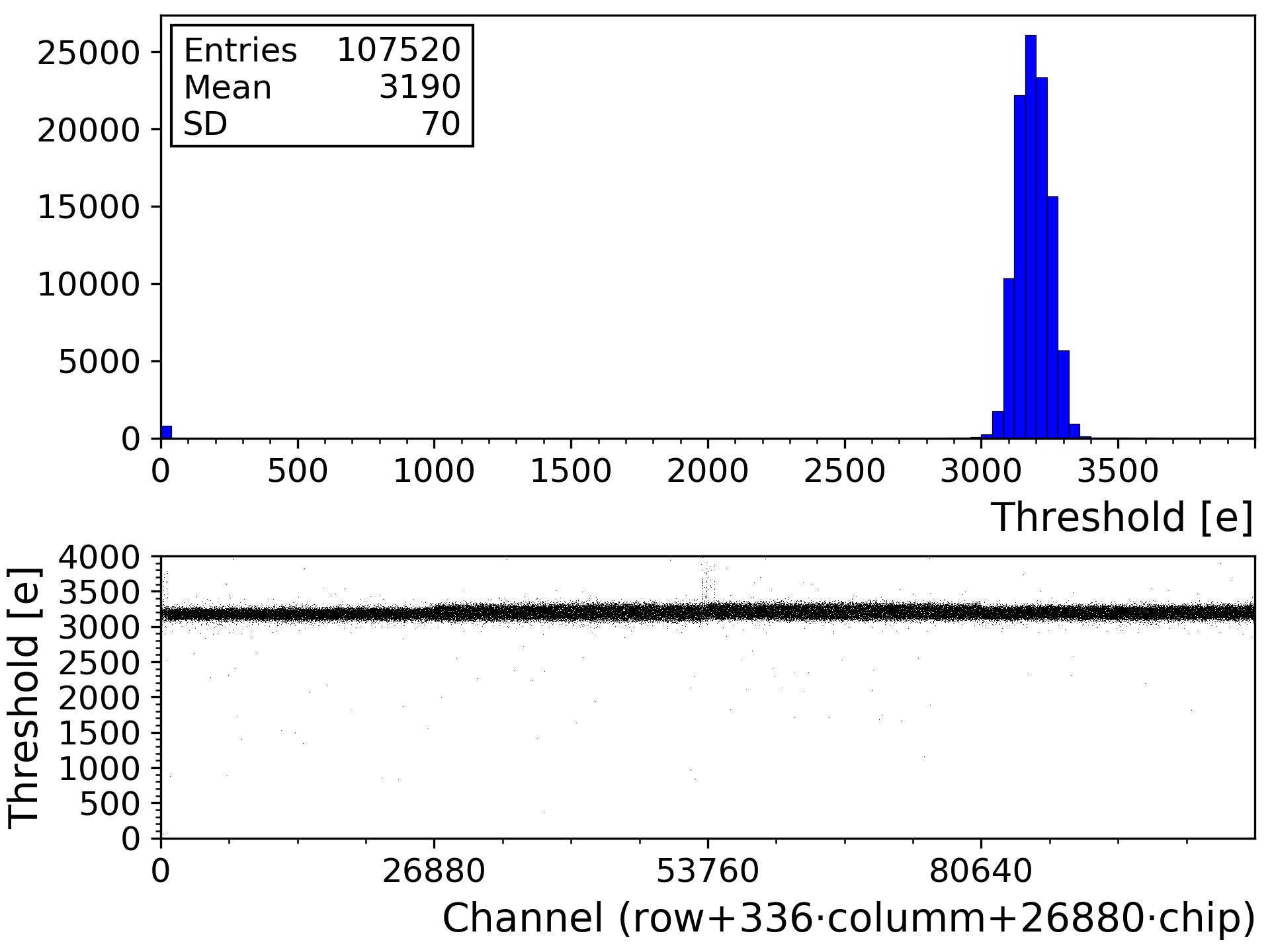}
\includegraphics[width=.49\textwidth]{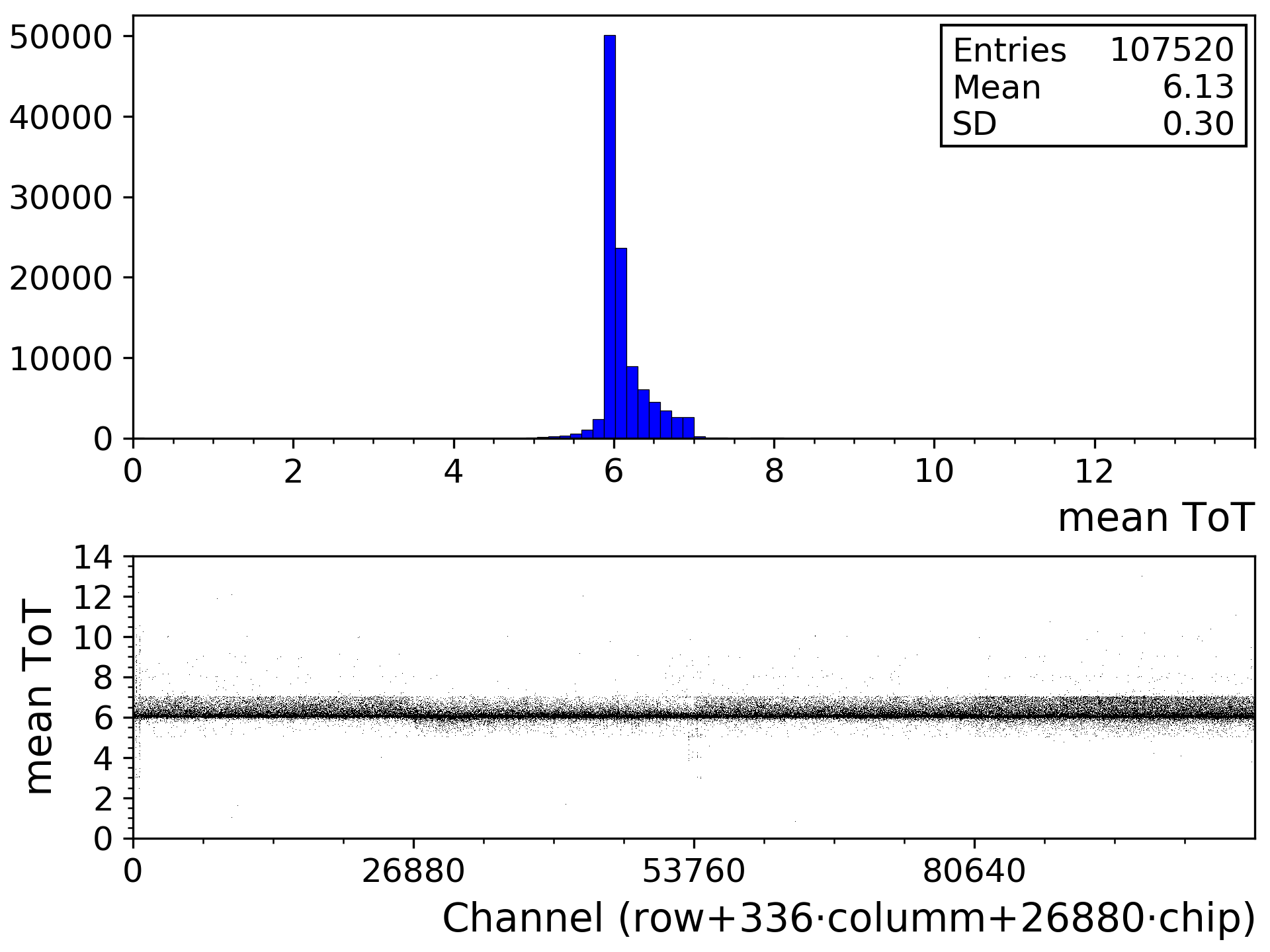}
\caption{\label{fig:tuning} Threshold and ToT distributions of an irradiated quad module after tuning.}
\end{figure}

Before irradiation, measurements were performed using a tuning with a threshold of \SI{$(3182 \pm 65)$}{e} and a response of \SI{$(6.00 \pm 0.09)$}{ToT} units at a reference charge of \SI{20}{ke}.
The results for the tuning used for measurements after irradiation are a threshold of \SI{$(3190 \pm 70)$}{e} and a response of \SI{($6.1 \pm 0.3$)}{ToT} units at a reference charge of \SI{20}{ke}. These distributions are shown in \Fig{fig:tuning}.

\section{Source measurements}
\begin{figure}[tbp]
        \centering 
        \includegraphics[width=.70\textwidth]{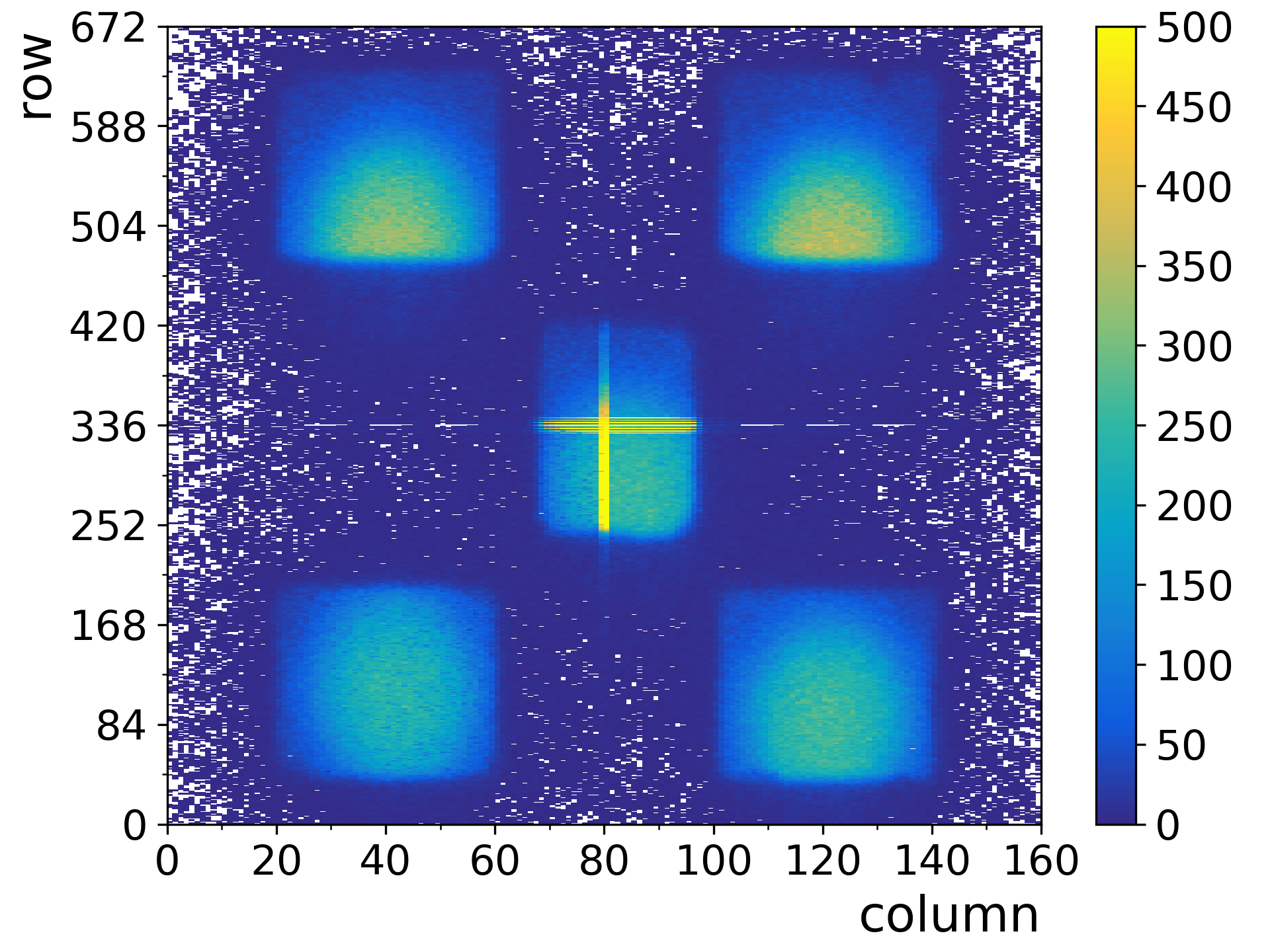}
        \caption{\label{fig:hitmap} Combined hit map of five different source positions on the unirradiated quad module. PCB openings are clearly visible as rectangular shapes. `Ganged' and `long' pixels have an increased hit count due to their increased area.}
    \end{figure}

	Before irradiation, measurements were performed with a Sr-90 source. Emitted $\beta$-particles pass through the sensor, the read-out chip, the PCB or its openings and are finally detected in a scintillator, which sends a read-out trigger to the chip. This methodology reduces noise from scattered particles, but if a backscattered particle is passing through the sensor in coincidence with a later triggered particle, both deposited charges are measured.	
	Starting with the raw data, \textit{fei4Analyzer}\footnote{\url{https://github.com/terzo/fei4Analyzer}} is used to match hits in adjacent pixels to a hit cluster.		
	A combined hit map of five different source positions is shown in \Fig{fig:hitmap}.	
	Most particles passing though the PCB are stopped or scattered before reaching the trigger scintillator, but the beam spot is clearly visible in the PCB openings.	
	Apart from beam spots, pixels with increased hits correspond to pixels with increased area, i.e. the `ganged' and `long' pixels in the border area between the front-end chips. This is a purely geometric effect.
    
\begin{figure}[hbt]
	\begin{minipage}[t]{0.49\textwidth}
		\centering
		\includegraphics[width=0.94\textwidth]{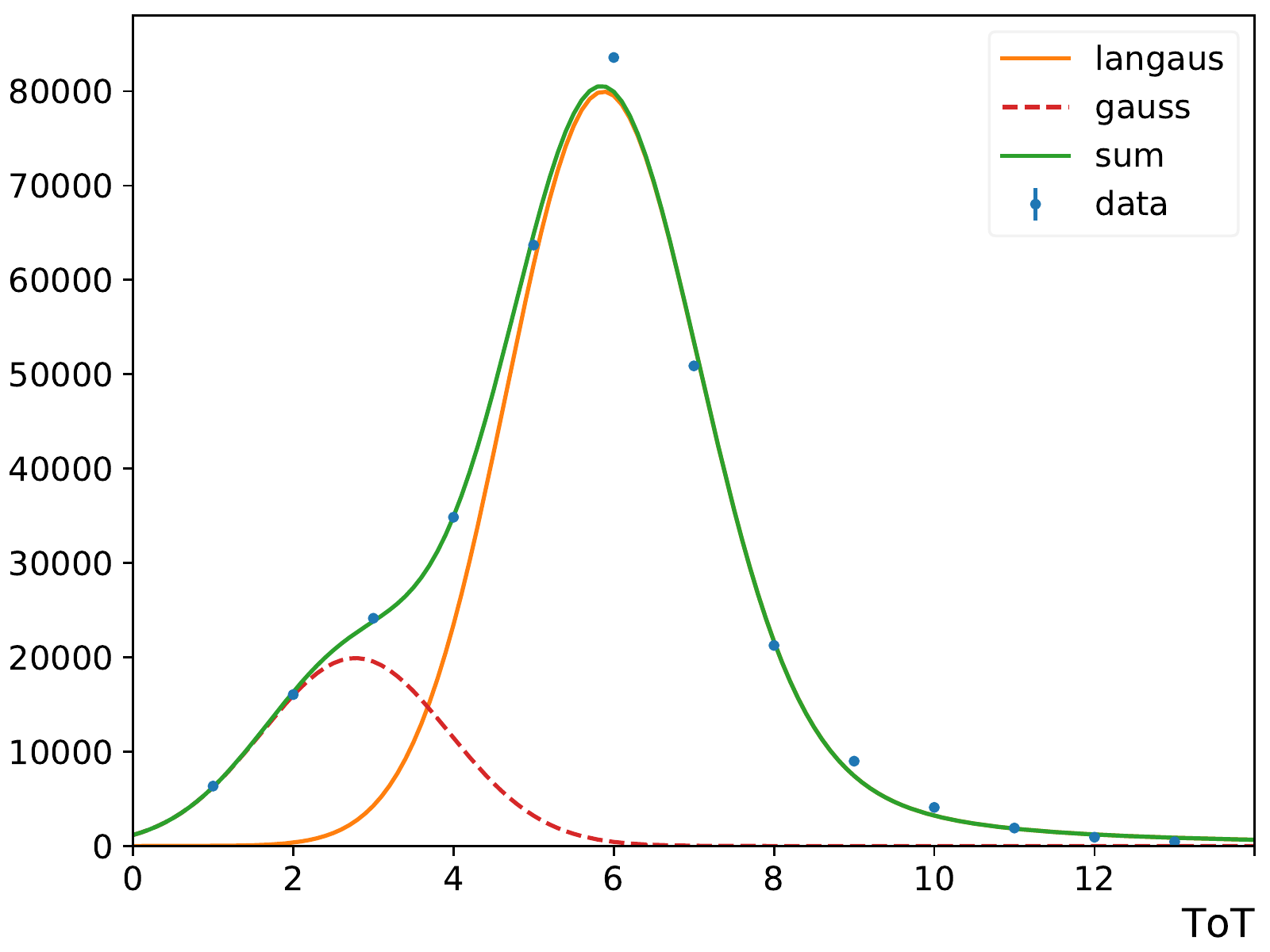}
		\caption{\label{fig:langauss+gauss} Collected ToT signal of the non-irradiated quad module for clusters of size 1, fitted with a Langauss distribution and additional Gaussian background.}
	\end{minipage}
	\hfill
	\begin{minipage}[t]{0.49\textwidth}
		\centering
	    \includegraphics[width=0.94\textwidth]{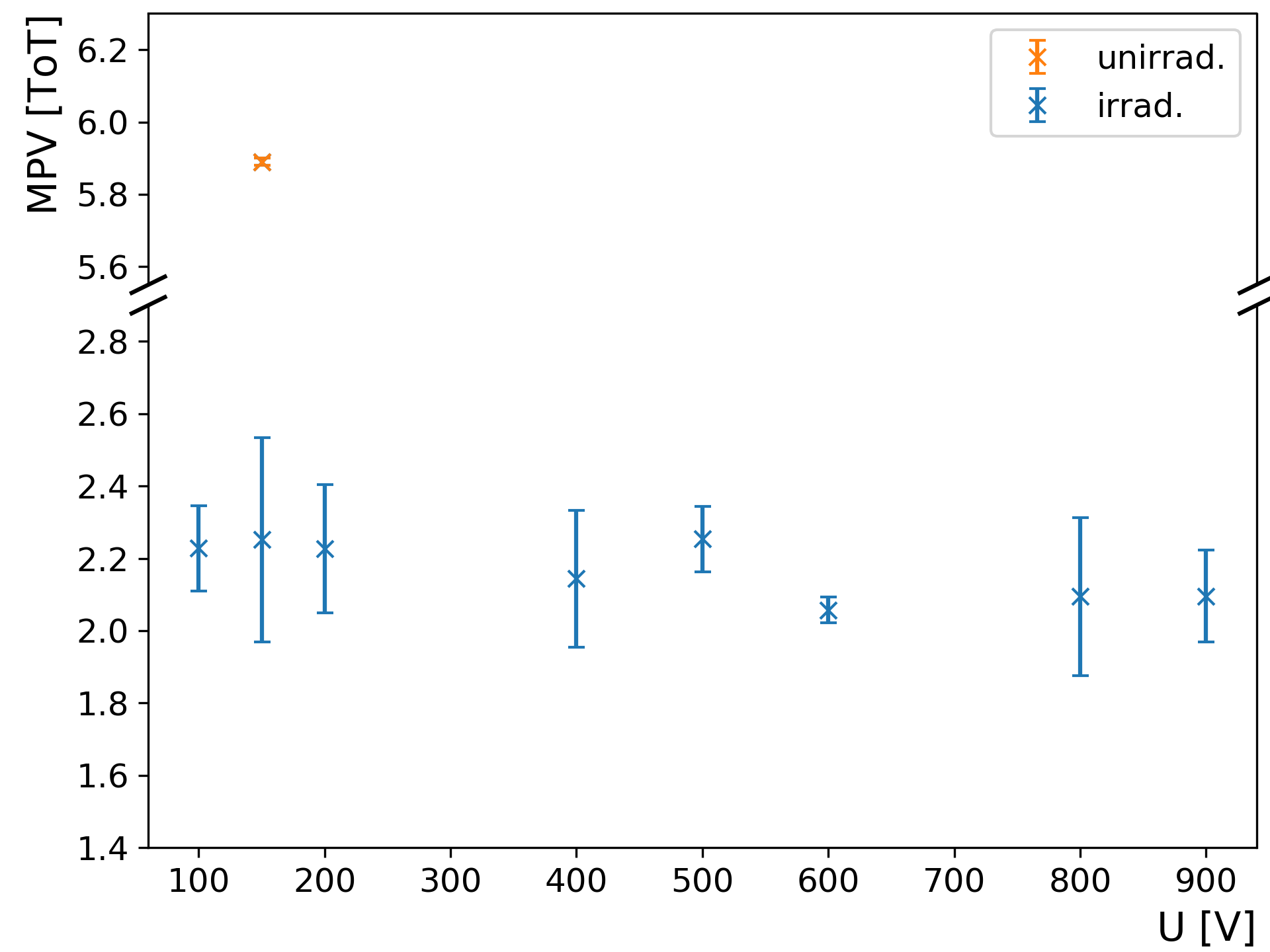}
	    \caption{\label{fig:charge_quad_irrad} Collected ToT signal vs. bias voltage of the irradiated quad module. The error bars result from the fit. The MPV obtained before irradiation is also shown.}
	\end{minipage}
	\end{figure}

These measurements were performed at a bias voltage of \SI{150}{V}.
The ToT information, which corresponds to the collected charge, was obtained from this data and histogramed, sorted by cluster size.
The resulting distribution for clusters of size 1 was fitted using a Landau-Gauss-convolution provided by \textit{pyLandau}.\footnote{\url{https://pypi.python.org/pypi/pyLandau}} To represent non-suppressed background from backscattered particles, a Gaussian is added.
    The fit for the unirradiated module is shown in \Fig{fig:langauss+gauss}. The fit result for the MPV of the Landau-Gauss-convolution is \SI{$(5.87 \pm 0.01)$}{ToT} units. 
    
    After irradiation, the module's own radiation caused by activation was measured, using FE self-triggering, since the $\beta$-source setup was not available. This method is more susceptible to noise. A bias scan was performed. The data was evaluated using the method described above.
    The result of MPV vs. bias voltage for the irradiated module is shown in \Fig{fig:charge_quad_irrad}. The error bars result from the fit covariance matrix. The average MPV is \SI{2.18}{ToT} units.
    As described in \autoref{sec:tuning}, the same tuning was chosen for all measurements.
    This comparability allows to determine a reduction of the ToT signal to \SI{40}{\percent} of the value before irradiation. This degradation can be improved easily by adjusting the tuning, leading to a larger ToT response at a similar charge in the sensor.    
    
 Due to the power dissipation of four FE chips, the external temperature sensors did not provide reliable values, but the sensor temperature was estimated to be \SI{-23}{\textdegree C} by comparing the leakage current during operation with the results obtained in \autoref{sec:IV}.    

\section{Testbeam measurements}
\begin{figure}[htbp]
        \centering 
        \includegraphics[width=.5\textwidth]{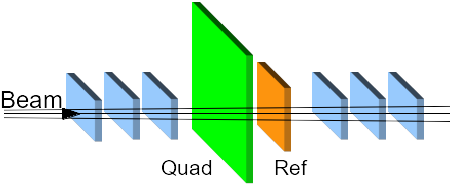}
        \caption{\label{fig:testbeam_sketch} Sketch of the testbeam setup at CERN-SPS, revealing the relative position of the quad and the reference to the Mimosa planes (blue).}
    \end{figure}

    Before irradiation, measurements were performed with a pion beam of \SI{120}{GeV} at CERN-SPS beamline H6. High tracking resolution is provided by six \mbox{MIMOSA26} sensors of ACONITE, an EUDET-type telescope \cite{Jansen2016}. 
    An unirradiated FE-I4 planar pixel module was used as the reference plane.
    A sketch of the setup is shown in \Fig{fig:testbeam_sketch}. A bias voltage of \SI{150}{\volt} was applied to the quad sensor during all measurements. Unless stated otherwise, its front-end chips were tuned to a threshold of \SI{3200}{e} and a response of \SI{6}{ToT} at a reference charge of \SI{20}{ke}.
    
\begin{figure}[htbp]
   	\begin{minipage}[t]{0.49\textwidth}
   		\centering
   		\includegraphics[width=01.0\textwidth]{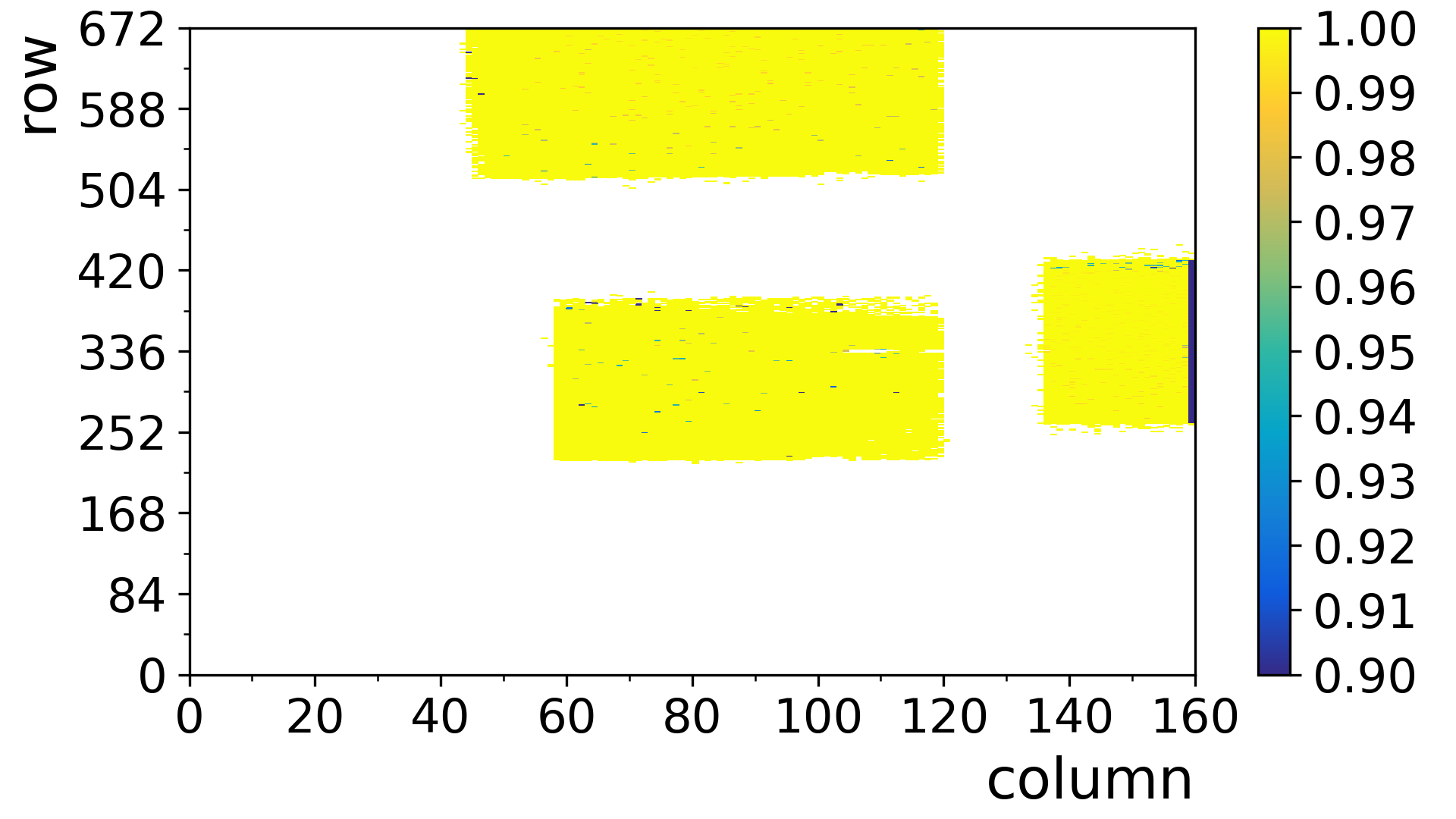}
   		\caption{\label{fig:eff_DUT20} Efficiency map for the complete sensor for three positions.}
   	\end{minipage}
   	\hfill
   	\begin{minipage}[t]{0.49\textwidth}
   		\centering
	    \includegraphics[width=01.0\textwidth]{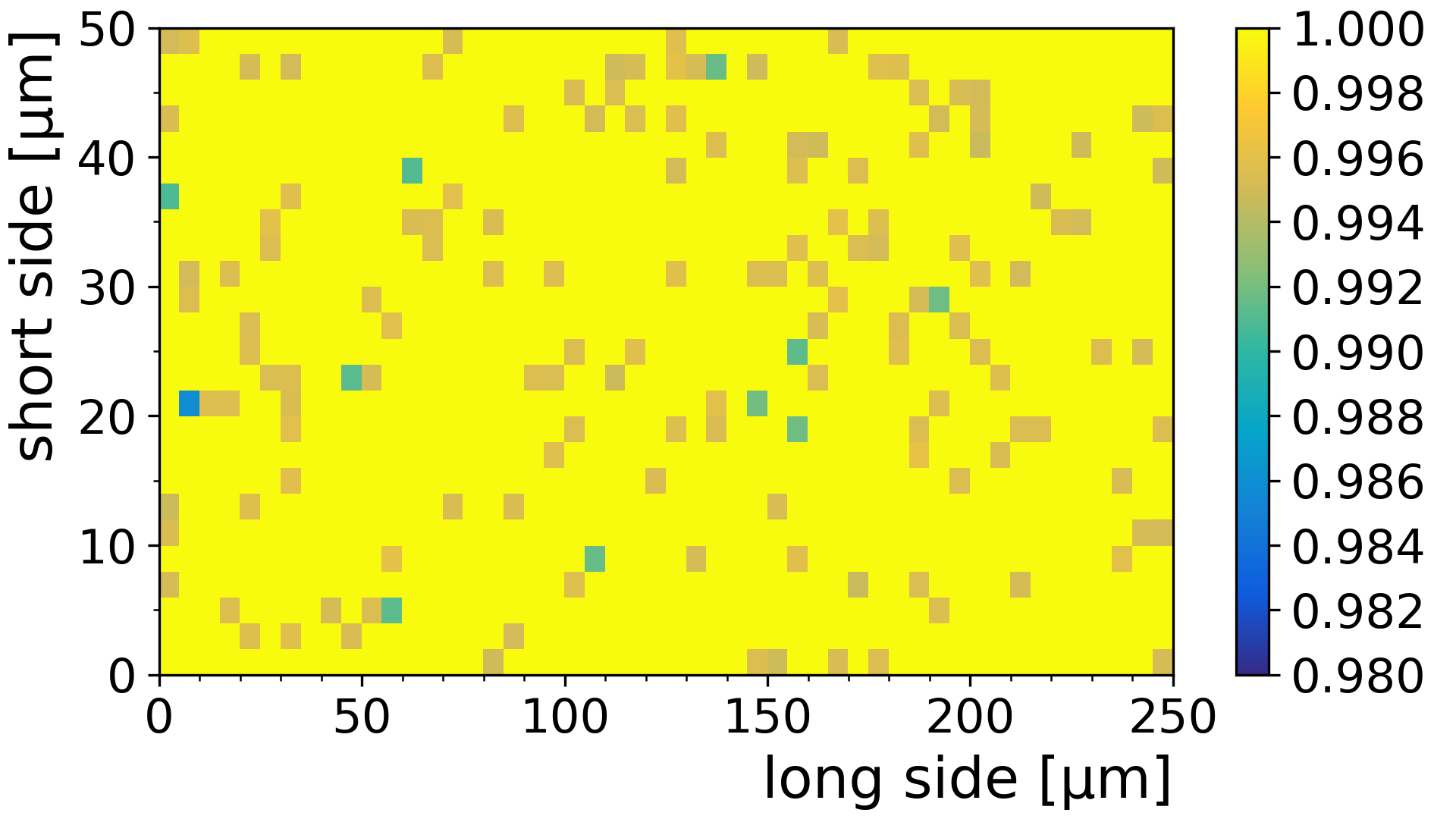}
	    \caption{\label{fig:eff_Geo2} In-pixel efficiency map for a standard pixel.}
   	\end{minipage}
   	\end{figure}

\begin{figure}[htbp]
   	\begin{minipage}[t]{0.49\textwidth}
   		\centering
   		\includegraphics[width=01.0\textwidth]{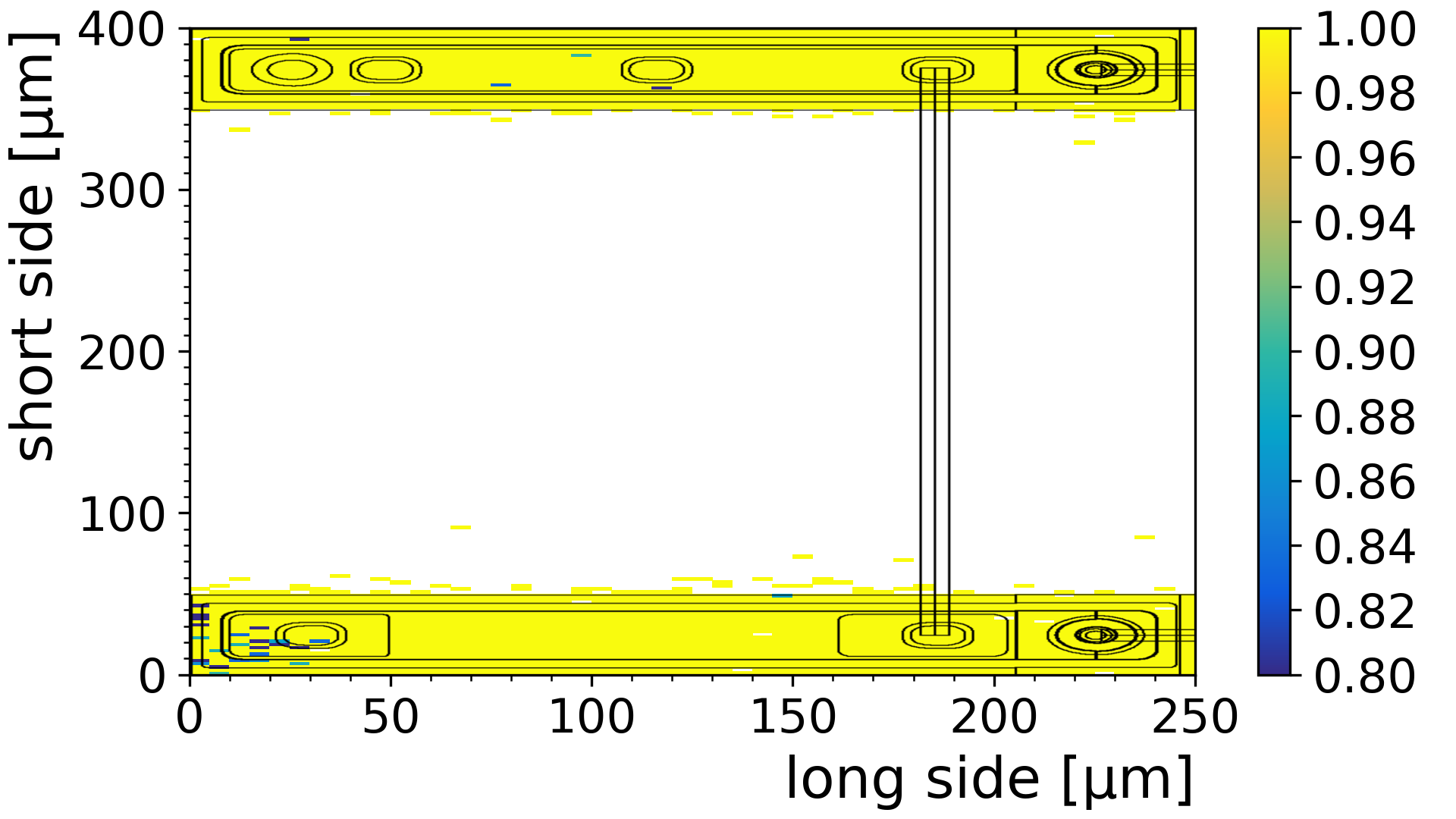}
   		\caption{\label{fig:eff_Geo47} In-pixel efficiency map for a ganged pixel, overlayed with its layout.}
   	\end{minipage}
   	\hfill
   	\begin{minipage}[t]{0.49\textwidth}
   		\centering
	    \includegraphics[width=01.0\textwidth]{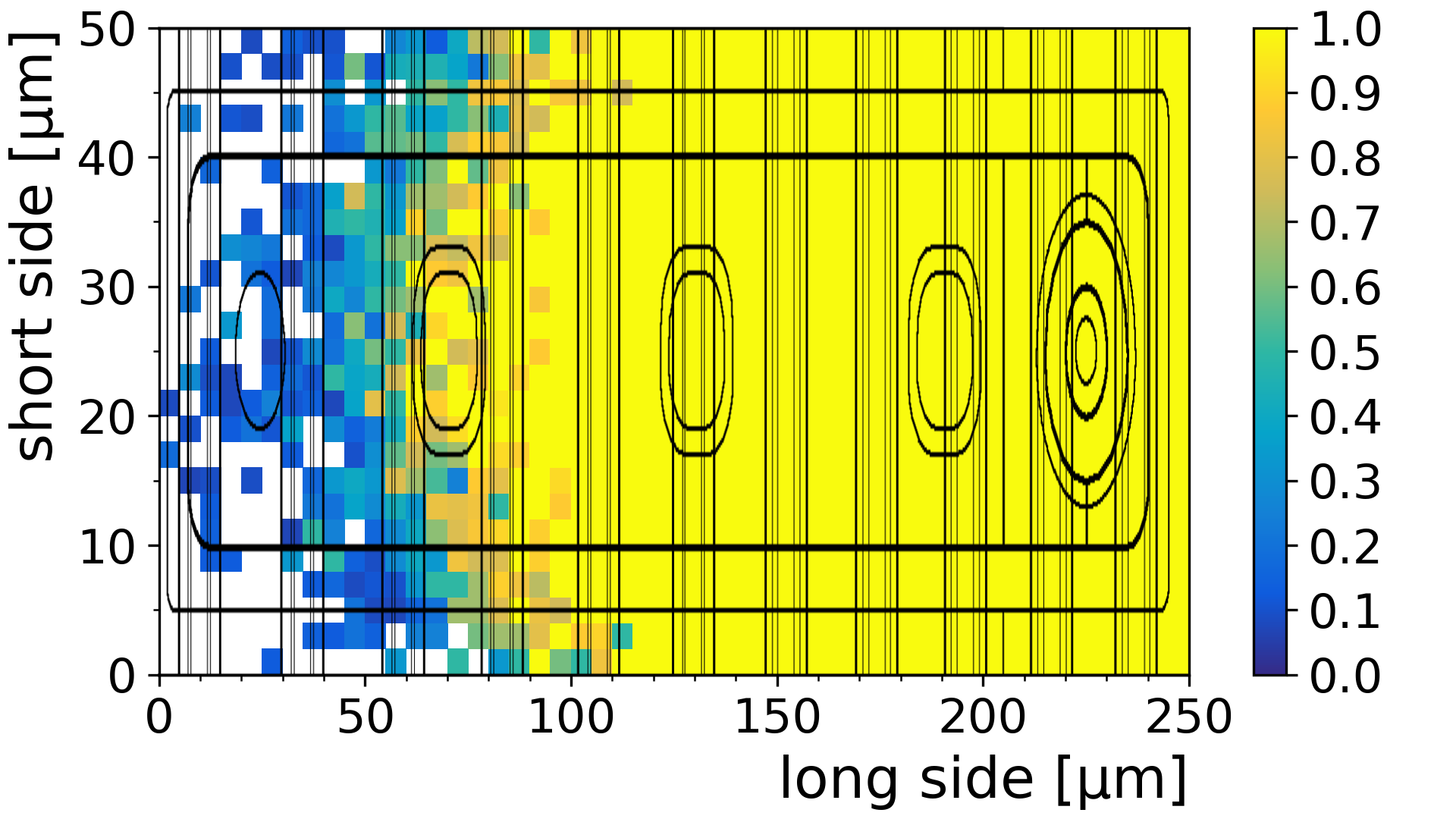}
	    \caption{\label{fig:eff_Geo3} In-pixel efficiency map for an edge pixel, overlayed with its and the guard ring layout.}
   	\end{minipage}
   	\end{figure}

    Track reconstruction was performed with \textit{EUTelescope}.\footnote{\url{http://eutelescope.web.cern.ch/}}
	For timing reasons, only tracks given by the telescope which match to hits in the reference plane are considered in the number of tracks $n_\text{tracks}$. Each track which is also matched with a hit in the quad is considered in the number of hits $n_\text{hits}$. 	
	The pixel matching margin was set to \SI{125}{\micro\meter} (\SI{50}{\micro\meter}) in X (Y),
	the cluster matching margin was set to \SI{400}{\micro\meter} (\SI{250}{\micro\meter}) in X (Y).	
	This analysis was performed with \textit{TBmon2}.\footnote{\url{https://bitbucket.org/TBmon2/tbmon2}}
	The efficiency $\varepsilon$ is defined as
	\begin{equation}
		\varepsilon = \frac{n_\text{hits}}{n_\text{tracks}},
	\end{equation}
	its relative error $\sigma_\varepsilon$ is defined as
	\begin{equation}
		\sigma_\varepsilon = \sqrt{\frac{\varepsilon \cdot (1 - \varepsilon)}{n_\text{tracks}}}.
	\end{equation}
	The recorded data were divided into runs with \SI{16}{k} events each. The efficiency and its error was calculated for every run and every pixel geometry. A weighted mean was determined for all runs taken under the same condition.	
	\Fig{fig:eff_DUT20} shows the combined efficiency map for three different measurement positions. No distinction can be observed between the `standard', `long', `ganged' or `inter-ganged' pixel designs. 
	\Fig{fig:eff_Geo2} shows the in-pixel efficiency map of the `standard' pixel design. The resulting efficiency is \SI{$(99.94 \pm 0.04)$}{\percent}.
	\Fig{fig:eff_Geo47} shows the in-pixel efficiency map of one `ganged' pixel design. The resulting efficiency is \SI{$(99.8 \pm 0.6)$}{\percent}.
	It is overlayed with the layout of the two pixels with a distance of \SI{300}{\micro\meter}, connected via a metal trace. The layout of the pixel in-between is not drawn.
	Comparable results with efficiencies well above \SI{99.5}{\percent} were obtained for all other `ganged/inter-ganged pixels' as well as `long pixels'.  
\Fig{fig:eff_Geo3} shows the in-pixel efficiency map of the `edge' pixel design. The resulting efficiency is \SI{$(76 \pm 6)$}{\percent}, but it is revealed that the sensor is full efficient up to \SI{150}{\micro\meter} beneath the guard rings.	 
	By lowering the threshold to \SI{1600}{e} an improvement could be reached. The resulting efficiency of the `edge pixel' for this threshold is \SI{$(84 \pm 6)$}{\percent}, the sensor is fully efficient up to \SI{200}{\micro\meter} beneath the guard rings. 	
	This is consistent with results of IBL pre-\mbox{studies \cite{Wittig2013}}, where \SI{500}{\micro\meter} long pixels were partially shifted beneath the guard rings.

\section{Summary}
Fully working n-in-n quad modules were assembled. Promising results have been obtained for these modules in laboratory and testbeam measurements. MIP-like particles deposit a charge signal corresponding to \SI{20}{ke} at a threshold of \SI{3200}{e}. A tracking efficiency of more than \SI{99.9}{\percent} has been measured for pixels with standard design which contribute to \SI{95}{\percent} of all read-out channels. 
A successful analysis of pixel efficiency for `ganged', `inter-ganged' and `long' designs was also performed, revealing a tracking efficiency well above \SI{99.5}{\percent} for these special read-out channels.

After a first irradiation step, a slight degradation is visible as expected: The leakage current increases and the collected charge decreases, but no failures have been observed.
Testbeam results after the next irradiation step are expected in October, followed by extensive lab investigations.


\acknowledgments
%

The possibility to irradiate detector samples at the CERN Proton Synchrotron Radiation
Test Facility IRRAD \cite{Ravotti:2014} is kindly acknowledged, especially the help by F. Ravotti and G. Pezzullo. 

Special thanks go to K. Wraight for the organization of this irradiation campaign, for the opportunity, and for the support to perform measurements with the irradiated quad module at the University of Glasgow.

The work presented here is carried out within the framework of Forschungsschwerpunkt FSP103 and supported by
the Bundesministerium f\"ur Bildung und Forschung BMBF under Grants 05H15PECAA and 05H15PECA9.

%
%
%
%
%
%
%
%
\bibliography{bib-psd11}
\bibliographystyle{JHEP}
\end{document}